# Formal Model of Living Systems

Margareta Segerståhl

Department of Biomedical Engineering and Computational Science (BECS)
P. O. Box 12200
Aalto University, School of Science
FI-00076 Aalto, Espoo, Finland
margareta.segerstahl@aalto.fi

**Abstract**

Here, a conceptual modeling formalism is proposed for the description and study of living and life-like systems. It is based on the real life and evolution of biological organisms and ultimately the system framework is determined by two functions, reproduction and survival – the two main fitness components of natural life. In the simplest form the model is a formal description of a discrete reproduction-survival state-transition system. The initial structure of the model can evolve to become more complex and it holds inherent potential for producing numerous variations of the basic theme. It is proposed, that this modeling formalism provides abstract system basis and immediately applicable conceptual tools for holistic study of living organisms on multiple levels of organization, and thereby for the understanding life and organization of complex living systems from a common systems point of view. The modeling principle is very generic and may therefore be applied directly also to the study of engineered and artificial systems.

## Systems View

The concept of an organism as a prototypic "living thing" is very intuitive for humans. But understanding life in formal terms remains a great challenge for modern science. The immense diversity and complexity of biological organisms has made it difficult to find general answers as to what the "laws of living systems" may be and how they relate to the laws of physics and chemistry. Currently there is neither commonly approved scientific vocabulary nor practical conceptual tools for forming uniform descriptions of life and living systems.

For a long time, the development of a general formalism was perhaps not considered to be an absolute necessity in biosciences. Research progressed very well also in its absence resulting in great scientific and technological breakthroughs. In the modern world, however, the concept of life is extending with increasing force to different kinds of human-made realms. Examples include topics such as artificial life, robotics, engineered minimal cells and the design of nanoscale biomolecular production plants; There are algorithmic information-based evolving systems operating in commerce, traffic, and business, only to mention some. Researchers in many different fields currently find themselves pondering what is the essence of life and living. The question is: How do artificial and engineered systems and their properties relate to the prototypic biological organisms and natural life? How to test their "life-likeness" and living potential? How to predict the evolutionary trajectories of organism-like or life-like innovations that may reside outside the biochemistry-based internal world of cells and organisms, or even be completely information-based?

## Lists of Life

Currently, a typical way to define life is to form a list of properties that a system should have in order to be classified as living one. For example, Farmer and Belin (1990, 1991) proposed the following list of properties admitting at the same time that any such list is bound to be both imprecise and incomplete:
1. Life is a pattern in space-time, rather than a specific material object.
2. Self-reproduction.
3. Information-storage of self-representation.
4. A metabolism which converts matter and energy from the environment into the pattern and activities of the organism.
5. Functional interactions with the environment.
6. Interdependence of parts.
7. Stability under perturbations and insensitivity to small changes, allowing the organism to preserve its form and continue to function in a noisy environment.
8. The ability to evolve.

Another list (Koshland, 2002) provides seven key principles as basis for life:
1. Program.
2. Improvisation.
3. Compartmentalization.
4. Energy.
5. Regeneration.
6. Adaptability.
7. Seclusion.

These two examples already demonstrate that terminology is neither universal nor self-explanatory: No two terms are the same, yet some clearly point to the same direction. These kinds of lists also need to be accompanied by a discussion on how to interpret them and to apply them to different situations.

Overall, the lists reveal what might be the essence of the problem in terms of describing life and living systems. They are clearly complementary rather than competitive in nature. This can be seen to reflect the difficulty of making a distinction between essential and derived properties when it comes to defining the concept of life. The current situation is therefore, that *not only is life itself an enormously complex phenomenon, but also the many ways in which it is being defined and discussed in the scientific discourse* add an extra layer of conceptual complexity to the problem of dealing with it. This is far from the precision by which physical and chemical concepts can be used to describe and define phenomena in their respective fields.

Here, I take these contemplations further and propose a modeling approach that can be used to determine, define, and combine attributes of life in a more precise and orderly manner.

## Minimal Model Approach

The life and structure of living organisms can be addressed on multiple levels of organization, from molecules to ecosystems. Therefore, the task of choosing an appropriate level of observation is of central importance here – as always when modeling complex phenomena (Checkland, 2000, Flood and Carson, 1993). My solution is a two-step approach: First, only unicellular life is considered. The reason is that despite all the diversity and complexity of biological organisms, their basic constituent unit is the cell. Real unicellular organisms also demonstrate that a single cell can also be an entire organism. This starting point reduces the initial complexity of the modeling problem providing a rather precise entry point to the overall challenge of modeling organisms in general.

Then, by comparing the resulting minimal model view of unicellular life to what is generally known about real, more complex forms of cellular life, including the structure, function, and evolution of different kinds of life cycles and multicellular organisms (plants, animals, and fungi), it is possible to propose how they too can be modeled within the same principle formalism.

## System Outline

In the study of life and living organisms it is typical to focus on mature forms of actively living cells and organisms when they use energy to perform all kinds of functions. For example they can grow, move, produce many kinds of metabolites, respond to stimuli, interact and reproduce. It is also typical to assume that if an organism cannot perform these activities, it will die. However, many kinds of unicellular organisms clearly demonstrate that is not necessarily the case. Although the metabolically active reproducing form is the one that is usually studied, these organisms are often able to alter their appearance completely in order to adopt an alternative form of existence as some kind passive inert survival structures. An example of this could be the formation of bacterial spores (Morita 1990), fungal spores, or seeds of flowering plants.

Cells of real unicellular organisms do not appear to abandon active state haphazardly. Instead the required cell-developmental changes and events are a response elicited by environmental clues that signal imminent or immediate energy deprivation (see for example the introduction in Hadany and Otto, 2007). It is typical, that once the transformation is complete the organism can withstand extreme conditions that would be destructive to the actively living form. The spore germinates when the environment again improves, and the organism returns to the active state of existence.

Based on this I propose the following minimal formalism for describing unicellular life (summarized in Figure 1). There are two states in which the conceptual living system (cell or organism) can exist. When conditions are favorable, the organism is in the *active state* and it comprises all the typical functions of metabolically active cellular life. *Reproduction* is considered to be an output function of active state living. Maintaining this state requires energy and the cell must obtain it from the environment or from its own internal energy stores. If the cell runs out of energy while it is in the active state, it will die.

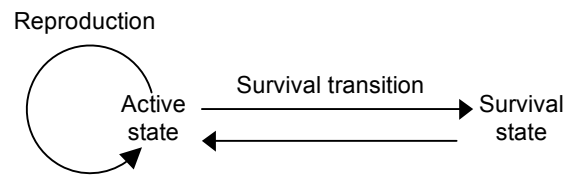

Figure 1: Minimal model formalism of living systems, based on unicellular life of real biological organisms. There are two states in which the system can exist. In the active state the system can use energy and nutrients for metabolically active living, growth, and reproduction. Formally the active state has two alternative system outputs: Reproduction (circular arrow) replicates the system without altering its state whereas survival transition takes the system into survival state. Both events require some amount of time and energy to be complete. The organism's state transition dynamics are regulated by energy availability.

The system can perform a *survival transition* when its environment turns hostile. This takes it from the active state to *survival state*. Real unicellular organisms must undergo changes in their appearance and behavior and gene expression profiles must be altered to complete this state-change successfully. In accordance with this the formal transition process assigned to require some amount of time and energy to be complete. Any additional time and energy that the system may need for returning to the active state after entering survival state simply add to the initial cost of initiating a survival transition in the first place. The minimal model survival transition is assigned to be discrete and irreversible: Once initiated, the system's only viable alternative is to enter survival state successfully.

The survival state encloses the organism's defining information content, protecting and conserving it. The survival state can be maintained without using energy and theoretically it provides indefinite passive existence for the living system in question. The survival state is named after the only system function it provides for the organism, which is mere survival in the sense that the defining information of the organism remains in existence, albeit in an inactive form for the time being.

## The Opposite Ends of Life

In this systems formalism reproduction and survival functions define two ultimately opposite alternatives for the way in which a living organism can remain in existence. An organism that relies entirely only on reproductive active-state living corresponds to a biphasic system case where the probability for entering survival state is zero. A hypothetical example is a unicellular marine alga that exhibits no survival state in its life cycle but instead, all cells are metabolically active reproducing entities that die if active living becomes impossible. This kind of life portrays a probabilistic metapopulation-type (Fronhofer et al. 2012) life-history strategy as a combination of efficient reproduction, strong-enough dispersal of actively living cells, and a satisfactory abundance of suitable free niches at any given moment in time that the organism can potentially reach and inhabit.

An extreme case of the opposite type could be presented by an organism that has demonstrably lived, subsequently entered survival state, and then appears to remain there indefinitely thereby approaching the borderline of even being a living system anymore. A real life counterpart could be a bacterial spore or a dry seed of a flowering plant, both being what may be called individuals (of which the latter is a multicellular entity). If these structures do not exhibit any sign of active life, are they then alive or dead, living or inanimate at the time of observation?

Ultimately the "aliveness" of the organism in a situation like this cannot be determined unless we try to germinate the seed or spore. Depending on the species these kinds of structures can be extremely stable and inert, but not all of them will germinate. Not knowing beforehand what the outcome of the germination attempt would be for any one unit structure (success or failure) we may have to consider that in its survival form proper it is both alive and dead at the same time, in which case this survival stage in the system's life presents a quantum state to the conceptual systems formalism of living system dynamics.

Interestingly, the existence of formal possibility for extremely stable survival-state occupancy makes it look like almost any kind of structural entity could be in this state and examined in the light of this living systems formalism. This could raise the issue of whether the proposed model outline is even too general, but the answer is no. It is correct that the survival state given, this model of living systems may also accommodate structurally organized states of inanimate matter. *Therefore*, being a living system comes down to the properties of the structural constitution of the individual living entity in question, its formal information storage properties, and the specific information that it contains. These features together specify the extent to which a structural organization can exhibit live behavior as determined by the biphasic reproduction-survival state transition framework.

All this results in an extended formal view of life as a system that on one side is conceptually determined purely by the survival function where it is conceptually allowed to apply also to inanimate organizations. On the ultimately opposite end of life resides pure and utter self-replication of the information-containing structure in question. This functional end provides a positive feedback loop for multiplicative existence of structural information-containing entities and possibly for their sustainable existence, formally for any kind of organization that can perform this function at a rate that is higher that the dissociation rate of the structures produced.

Not all information-containing structure in all environments have properties that enable them to live. But some clearly do, the biological organism being a prototypic example. A usual assumption in biological research is that organisms evolve to maximize reproduction. But in this model this evolutionary tendency is inherently embedded in the model formalism: There will simply be more cells that evolve towards maximizing reproductive active state living, than cells that spend excessive time in their survival state.

## Evolution In-between Reproduction and Survival

The ability to undergo adaptive evolution is a defining feature of biological organisms. Starting from the minimal-model case, the formal reproduction-survival state-transition framework may adapt and evolve to its environment in many different ways, just like a cell can undergo Darwinian adaptive phenotypic evolution with mutation and selection.

Adaptive periodic entry into survival state in a predictable fluctuating environment provides a mechanism for continuous long-term evolution towards more efficient active state living. When a cell returns from the survival state after a period of stress that would have otherwise killed it, it can continue its evolution recursively from the adaptational state where the previous round of active evolutionary living and reproduction had taken it. Sporadic entries into survival state are less likely to be equally effective in this sense, because they may take a cell into survival state also when it could alternatively undergo active state living. However, they can provide a probabilistic back-up system for organism's survival in the case of stochastic catastrophic events that can take an actively living cell completely by surprise.

In the simplest possible minimal model scenario the survival transition may be direct and discrete, but in reality it seldom is. It appears that the cells of most real organisms spend most of their time in states where they invest simultaneously and in varying ratios to both reproduction ability and survival probability. In the system formalism they occupy *intermediate metastable transition states* that reside formally in-between the ultimately theoretical system-defining end-point states of maximum reproduction and absolute survival (see Figure 2). Because the overall evolution of a reproduction-survival state-transition system is towards active-state living, these intermediate states are likely to emerge into survival transition biology in the course of evolution as opposed to more immediate system progression in the direction of entering survival state.

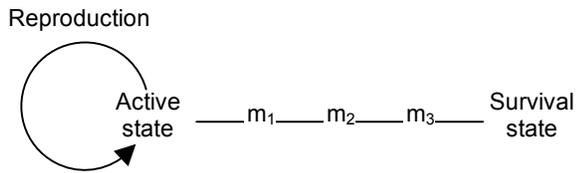

Figure 2. A simple description of intermediate metastable survival transition states (m-states) that position along the linear continuum of survival transition progression in the formal state-space of survival transition biology.

Unicellular organisms with complex life cycles can be considered from this same survival transition point of view. For example, when the yeast *Schizosaccharomyces cerevisiae* leaves the active state in order to form spores, it must proceed through several identifiable stages that provide structure to its survival transition biology. The life cycle of the unicellular malaria-causing parasite *Plasmodium falciparum* is very complex and in many respects entirely different from the yeast life cycle, but it too can be viewed from the point of view of survival transition biology. It can be seen as a stage-wise realization of a balancing act between cellular reproduction and survival functions. Flexible cell-type differentiation processes take the system from one stage to the other as this organism of reproducing cells migrates through different tissue types in its two host organisms: the mosquito and the human.

In colonial multicellularity, presented for example by some aquatic green algae, single-cell individuals form multicellular entities but each cell retains its individualistic identity and potential for future reproduction. This kind of investment to cellular properties that enable multiple cells to organize into a single functional structure can physically aid the survival of them all. Functions that allow this behavior are not directly related to the reproduction operations of a single cell, but on system level they may still contribute to this aim – especially if conditions are not ideal for single-cell living when the colony forms and the cells in question are likely to be eventually forced to enter survival state.

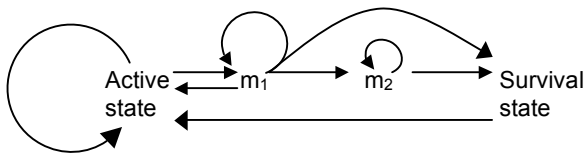

Figure 3. An arbitrary example of an evolved survival transition framework topology with metastable transition states $m_1$ and $m_2$. After leaving the active state the cells of this organism may still return directly from the $m_1$-state. Later each cell proceeds with the transition, either directly or via another intermediate state $m_2$ that might only be available for the cells of a supporting cell line that further protect those cells of the organism that transit directly from $m_1$ to survival state.

**Multicellular Organisms**

On one hand, complex unicellular life cycles can be formally considered to be patterns in the state-transition formalism. On the other hand, the different cell lineages of complex multicellular organisms can each be interpreted in the same way from the same state-transition perspective regarding *both* the evolution *and* the actual developmental progression of their cellular differentiation processes.

For example, rapidly proliferating cells of an early human embryo differentiate progressively as the organism matures, their reproduction diminishes. The cells acquire features that contribute to the integrity of the developmentally complex multicellular human entity, through cell-type specific functional and structural differentiation. They adopt a role as part of a multicell-complex (a human individual) as intermediate metastable survival transition cells, and instead of entering survival state they get to live in terms of their own higher-level organizational entity in places and environments where they could never live as solitary single-cell individuals. Uncontrolled reproduction is no longer the sought-for state for the primary existence of this organism's cells, as demonstrated by cancerous cells (Merlo et. al. 2006).

Multicellular individuals are considered to be organizational entities of lower-level unit-structures. As individuals, they become subject to the same system principles of reproduction-survival state-transition dynamics as are the cells on the lower level of organization. This adds a dimension to the overall complexity of the individual organism in question. The next level of complexity expectedly arises when the multicellular individuals form social groups. A starting point for addressing this kind of conceptual *living complexity* as a formal problem in the study of complex adaptive systems in general, is to propose that the more layers of living organization there are in the formal description of an individual organism, the more complex and multilayered the interaction patterns within and between reproduction-survival state-transition dynamics on all levels of its organization.

## Conclusion

A general scientific model of life must present it as something that can be established in the realm of chemistry, obeying the laws of physics, and being manifested in the form and function of all kinds of biological organisms. The proposed modeling approach enables introduction of general systems formalism for the description and definition of very different kinds of biological systems from a common systems point of view. It should be possible to take the formal system framework of the hereby proposed model and examine it together with the key attributes of life that were listed at the beginning of this article, to see how they can best fit together and thereby deliver a concise picture of living systems in general.

In the light of this conceptual framework, *living* may refer to the things that happen in terms of this framework. *Life* on the other hand is the entire identifiable emergent overall phenomenon that arises from the operation of the living systems. More specifically, the existence of life on Earth may be seen as patterns in space-time that emerge from the structure-based functioning of a subset of information-containing systems. Based on their organizational information-containing properties, they undergo operations in their prevailing environment that *channel available energy*

(Morowitz and Smith 2007) and matter into the processes of self-maintenance and self-reproduction.

The structure of the hereby-presented reproduction-survival systems model stems from what is generally known about simple living of biological organisms. Attributes of the model are very generic suggesting, that it will be possible to use it also for testing whether other kinds of system possess the kind of formal system properties that are needed, in order to have potential for reproduction-survival dynamics and adaptive evolutionary living.

Much work is to be done in order to examine the full potential of the proposed modeling formalism and the rules by which the proposed system dynamics can operate and evolve. But the ease, by which many different versions of natural life can simply be immediately positioned even to this rough prototypic model schematic, makes this approach seem very promising. Gaining a simple and useful general formalism for the definition, description and study of living and life-like system across disciplines would be a great improvement to the current situation.

## Acknowledgements

I would like to thank Jörkki J. Hyvönen and Michael Patra for important conversations regarding the main idea of this work. I also thank Kimmo Kaski for useful discussions and for providing the necessary facilities to carry out this research.

## References

Checkland, P. (2000) Soft systems methodology: A thirty year retrospective. *Sys. Res. Behav. Sci.* 17:S11–S58.

Farmer, J. D. and Belin, A. d'A. (1990). Artificial life. The coming evolution. *Santa Fe Institute working paper* 90-003

Farmer, J. D. and Belin, A. d'A. (1991). Artificial life: The coming evolution. In *Proceedings of the Second Conference on Artificial Life*, pages 815–383. After Spafford, E. H. (1995). Computer viruses as artificial life. In Langton, C. G., editor, *Artificial Life: An Overview*, pages 249–265. MIT Press, Cambridge, MA.

Flood, R. L. and Carson, E. R. (1993) *Dealing with Complexity: An Introduction to the Theory and Application of Systems Science.* Plenum Press, NY. 2nd edition.

Fronhofer et al. (2012). Why are metapopulations so rare? *Ecology.* 93(8):1967–1978.

Hadany, L. and Otto, S. (2007). The evolution of conditions-dependent Sex in the face of high costs. Genetics 176(3): 1713–1727.

Koshland, D. E. Jr. (2002). The seven pillars of life. *Science* 295(5563):2215–2216.

Merlo, L. et al. (2006). Cancer as an evolutionary and ecological process. *Nat. Rev. Cancer* 6: 924–935.

Morita, R. (1990) The starvation-survival state of microorganisms in nature and its relationship to the bioavailable Energy. *Experientia.* 46:813–817.

Morowitz, H. and Smith, E. (2007), Energy flow and the organization of life. *Complexity*, 13: 51–59.